\documentclass[10pt]{bmc_article}

\usepackage{cite} 
\usepackage{url}  
\usepackage{ifthen}  
\usepackage{multicol}   
\usepackage[utf8]{inputenc} 
\usepackage{graphicx}

\newboolean{publ}

\newenvironment{bmcformat}{\fussy\setboolean{publ}{true}}{\fussy}

\begin{document}
\begin{bmcformat}

\title{Sequence alignment and mutual information}

\author{Orion Penner\correspondingauthor$^1$
       \email{Orion Penner\correspondingauthor - openner@phas.ucalgary.ca}
      \and
         Peter Grassberger\correspondingauthor$^{1,2}$
         \email{Peter Grassberger\correspondingauthor - pgrassbe@ucalgary.ca}
       and 
         Maya Paczuski$^1$
         \email{Maya Paczuski - maya.paczuski@ucalgary.ca}
      }

\address{
    \iid(1)Complexity Science Group, Department of Physics and Astronomy, University of Calgary, Calgary, Alberta T2N 1N4, Canada\\
    \iid(2)Institute for Biocomplexity and Informatics, Department of Biological Sciences, University of Calgary, Calgary, Alberta T2N 1N4, Canada
}

\maketitle

\begin{abstract}
        \paragraph*{Background:} Alignment of biological sequences such as DNA, RNA or proteins is one of 
the most widely used tools in computational bioscience.  All existing alignment algorithms rely 
on heuristic scoring schemes based on biological expertise.  Therefore, these algorithms do not 
provide model independent and objective measures for how similar two (or more) sequences actually 
are.  Although information theory provides such a similarity measure -- the mutual information 
(MI) -- previous attempts to connect sequence alignment and information theory
have not produced realistic estimates for the MI from a given alignment.
      
        \paragraph*{Results:} Here we describe a 
simple and flexible approach to get robust estimates of MI from {\it global} 
alignments. For mammalian mitochondrial DNA, our approach gives pairwise MI estimates for commonly 
used global alignment algorithms that are strikingly close to estimates obtained by an entirely 
unrelated approach -- concatenating and zipping the sequences. 

        \paragraph*{Conclusions:} This remarkable consistency may help establish MI as a reliable tool for evaluating the quality of global alignments, judging the relative merits of different alignment algorithms, 
and estimating the significance of specific alignments.  We expect that our approach can 
be extended to establish further connections between information theory and sequence alignment, 
including applications to local and multiple alignment procedures.

\end{abstract}


\section*{Background}
Sequence alignment achieves many purposes and comes in several different
varieties \cite{aluru2006}: Local versus global (and even
``glocal'':~\cite{brudno2003gaf}), pairwise versus multiple, and DNA/RNA versus
proteins. Rather than listing all applications, we cite just two numbers: The
two original papers on the BLAST algorithm for local alignment by \cite{altschul1990} 
and on one of its improvements \cite{altschul1997} have been cited more than 
43,000 times, and the number of daily file uploads to the NCBI server providing BLAST 
is $\approx 140,000$ \cite{mcginnis2004bcp}. A partial list of alignment tools 
in the public domain can be found in {\sf http://pbil.univ-lyon1.fr/alignment.html}.\pb

In {\it global} alignment, which we focus on here, two sequences of comparable length 
are placed one below the other.  The algorithm inserts blanks in each of the sequences 
such that the number of positions at which the two sequences agree is maximized.
More precisely, a {\it scoring scheme} is used where each position at which the
two sequences agree is rewarded by a positive score, while each disagreement
(``mutation'') and each insertion of a blank (``gap'') is punished by a
negative one.  The best alignment is that with the highest total score.  In
{\it local} alignment, one aligns only subsequences against each other and
looks for the highest scores between any pairs of subsequences. Regions that
cannot be well-aligned are simply ignored.  Existing codes use either heuristic
scoring schemes or scores derived from explicit probabilistic
models \cite{durbin1998bsa}.\pb

In either case, the absolute value of the score cannot be used to judge
reliably the quality or the significance of an alignment. As a result,
significance is typically estimated by aligning random sequences
(``surrogates'') and comparing the distribution of scores between these
surrogates to the scores between the true biological sequences. Significance
estimates are particularly relevant when aligning a sequence of interest
against an entire data bank, in order to find a homologue. In that case wrong
estimates for the tail of the distribution of pairwise ``similarities''  could
render the results worthless.\pb

In this context -- and in many others -- an objective measure for similarity
between two biological sequences would be extremely useful.  Such an objective
measure is provided by information theory \cite{cover}. Roughly, the {\it
complexity} $K(A)$ of a sequence $A$ is the minimal amount of information
(measured in bits) needed to specify $A$ uniquely. For two sequences $A$ and
$B$, the conditional complexity (or conditional information) $K(A|B)$ is the
information needed to specify $A$, if $B$ is already known. If $A$ and $B$ are
similar, this information might consist of a short list of changes needed to go
from $B$ to $A$, and $K(A|B)$ is small. If, on the other hand, $A$ and $B$ have
nothing in common, then knowing $B$ is useless and $K(A|B)=K(A)$. Finally, the
{\it mutual information} (MI) is defined as the difference $I(A;B) =
K(A)-K(A|B).$  It is the amount of information which is common to $A$ and
$B$, and is also equal to the amount of information in $B$ which is useful for
describing $A$, and {\it vice versa}. Indeed, it can be shown that, up to
correction terms that become negligible for long sequences (see~\cite{cover}): (a)
$I(A;B) \geq 0$; (b) $I(A;B) = 0$ if and only if $A$ and $B$ are completely
independent; (c)  $I(A;A) = K(A)$; and (d) $I(A;B) = I(B;A)$. Moreover, the
likelihood that  $A$ and $B$ arose independently is  $p = 2^{-I(A;B)}$
(see~\cite{milos95}).  Hence, the similarity is significant and not by chance
when $I(A;B)$ is large.\pb

The fact that alignment and information theory are closely related has been
realized repeatedly.  However, most work in this direction has focussed on
aligning images rather than sequences \cite{viola1997}. Conceptually, these
two problems are closely related, but technically, they are not. The effects 
of sequence randomness on the significance of alignments has also been
studied in \cite{Allison99}.  Finally, attempts to extend the notion of {\it edit
distance} \cite{aluru2006} to more general editing operations have been made.
In this case the similarity of two sequences is quantified by the complexity of
the edit string, see \cite{varre99}.  Indeed, the aims of \cite{varre99} are
similar to ours, but their approach differs in several key respects and leads to
markedly different results.

\section*{Methods}
\subsection*{Translation String}

\begin{figure}
\centering
\includegraphics[width=0.8\textwidth]{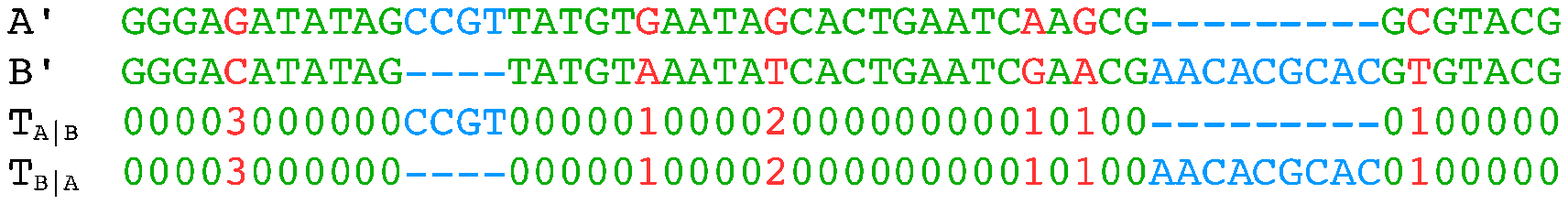}
\caption{Example of an alignment and of the two translation strings $T_{A|B}$
  and $T_{B|A}$. Colors indicate sites with mutations (red), gaps (blue), and 
  conservation (green).
  \label{fig:example_trans}}
\end{figure}

At the heart of our approach is the concept of a {\it translation string}.  The
{\it translation string} $T_{B|A}$ contains the information necessary to recover 
the sequence $B$ from another sequence $A$. Similarly, $T_{A|B}$ contains the 
information needed to obtain $A$ from $B$.  Here we focus on  DNA sequences,
consisting of the letters A,C,G and T, and  corresponding to complete
mitochondrial genomes.  But the approach is more general and can be applied  to
protein sequences without further effort.  We refer to the $i^{th}$ element of
sequence $X$ as $X_i$, and denote the length of $X$ as $n_X$.  Any global
alignment algorithm, when applied to $A$ and $B$, outputs a pair of sequences
$(A',B')$ of equal length $n\geq \max\{n_A,n_B\}$. The sequences $A'$ and $B'$
are obtained from $A$ and $B$ by inserting hyphens ("gaps") such that the total 
score is maximized. The strings $T_{B|A}$ and $T_{A|B}$ also have length $n$, and 
are composed from an alphabet of nine characters. For each $i$, the letter 
$T_{B|A,i}$ is a function of $A'_i$ and $B'_i$ only. An example of this
process is found in Figure~\ref{fig:example_trans}; the rules to create $T_{B|A}$
are as follows:
\begin{itemize}
\item if $A'_i = B'_i$, then $T_{B|A,i}=0$;
\item if $A'_i$ is a hyphen (gap), then $T_{B|A,i}$ has to specify explicitly what 
is in $B$; hence $T_{B|A,i} = B'_i \in\{$A,C,G,T$\}$;
\item if $B'_i$ is a hyphen (gap), then $T_{B|A,i}$ has to indicate that {\it something}
is deleted from $A'$, but there is no need to specify what. Hence $T_{B|A,i} = B'_i = -$;
\item if $A'_i \to B'_i$ is a {\it transition}, i.e. a substitution A$\leftrightarrow$G or C$\leftrightarrow$T, then $T_{B|A,i}=1$;
\item if $A'_i \to B'_i$ is a {\it transversion} A$\leftrightarrow$C or T$\leftrightarrow$G, then $T_{B|A,i}=2$;
\item if $A'_i \to B'_i$ is a transversion A$\leftrightarrow$T or G$\leftrightarrow$C, then $T_{B|A,i}=3$.
\end{itemize}
$T_{B|A}$ is defined such that $B'$ (and thus also $B$) is obtained uniquely from $A'$. 
But $A'$ can be obtained from $A$ using $T_{B|A}$. Thus $T_{B|A}$ does 
exactly what it was intended to do: it allows one to recover $B$ from $A$. It does not, 
however, allow one to recover $A$ from $B$. Due to the second and third bullet points
above, $T_{B|A}$ is not the same as $T_{A|B}$.  This distinguishes our
approach from typical edit string methods.

\subsection*{Mutual Information}

An estimate of the conditional complexity $K(B|A)$ is obtained by compressing
$T_{B|A}$ using any general purpose compression algorithm  such as zip, gzip,
bzip2, etc. In the results shown here we use lpaq1 \cite{mahoney}; see also this 
reference for a survey of public domain lossless compression algorithms). Denoting by 
${\rm comp}(A)$ the compressed version of $A$ and by ${\rm len}[A]$ the length of $A$
in bits, gives an exact upper bound
\begin{equation}
   K(B|A) \leq {\rm len}[{\rm comp}(T_{B|A})].                            \label{Kest}
\end{equation}
In order to obtain an estimate of  MI, we have to subtract $K(B|A)$ from $K(B)$, which is also 
estimated via compression. However, unlike $T_{B|A}$,  $B$ is a DNA string.
Since general purpose compression algorithms are known to be inferior for DNA \cite{gencomp,cao2007ssa}
we use an efficient DNA compressor called ``XM" \cite{cao2007ssa}.  The resulting MI estimate is
\begin{equation}
   I(A;B) \approx I(A;B)_{\rm align} = {\rm len}[{\rm XM}(B)] - {\rm len}[{\rm lpaq1}(T_{B|A})].
\end{equation}

At first sight it might seem paradoxical that $I(A;B)_{\rm align}$ can even be
positive.  Not only does $T_{B|A}$ involve a larger alphabet than $B$, but, in
general, it is also a longer string. Thus one could expect that $T_{B|A}$ would
not typically compress to a shorter size than $B$. The reason why this first
impression is wrong is clear from Figure~\ref{fig:example_trans}: If $A$ and $B$
are similar, then $T_{B|A}$ consists mostly of zeroes and compresses readily. In 
practical alignment schemes, the scores for mismatches are carefully chosen such 
that more frequent substitutions are punished less than unlikely substitutions. In
contrast, coding each mismatch simply by a letter in $T_{B|A}$ seems to ignore
this issue.  However, more frequent mismatches will give letters occurring with
higher frequency, and general purpose compression algorithms utilize frequency
differences to achieve higher compression.\pb

Conceptually our approach is similar to encoding of generalized edit strings
in \cite{varre99}. However, there are several pivotal differences between that
work and ours.  First, the authors in \cite{varre99} did not compress their
edit strings and as a result the conclusions they were able to draw from
a quantitative analysis were much weaker than ours.  Second, our approach
utilizes an alignment algorithm to achieve an efficient encoding of $T_{B|A}$.
In addition to producing a better estimate of $K(B|A)$, this allows us to make
quantitative evaluations of the algorithm itself.  An additional difference
between our approach and the traditional edit methods used in approximate
string matching \cite{navarro2001}  is that our {\it translation strings} do not
give both translations $A\to B$ and $B\to A$ from the same string.  This
asymmetry is crucial to establish the relations to conditional and mutual information.\pb

For long strings, $I(A;B)$ should be symmetric in its
arguments. In general, the estimates satisfy $I(A;B)_{\rm align}\approx I(B;A)_{\rm align}$
(see Figure~S3 in the supplementary material).
Indeed, the translation strings $T_{B|A}$ and $T_{A|B}$
can differ substantially, resulting  in different estimates for
$K(B|A)$ and $K(A|B)$ via Eq.~(\ref{Kest}). This
difference is mostly cancelled by differences between ${\rm
  len}[{\rm XM}(B)]$ and ${\rm len}[{\rm XM}(A)]$. Take, for instance, the case
where $B$ is much shorter than $A$.  Then $T_{B|A}$ 
consists mostly of hyphens and is highly compressible. On the other hand, $T_{A|B}$
is similar to $A$, since most letters have
to be inserted when translating $B$ to $A$. Thus both $I(A;B)_{\rm
  align}$ and $I(B;A)_{\rm align}$ are small compared to $K(A)$, but for 
different reasons. Further details are given in the supplementary material.

\subsection*{Tools}
We utilized the MAVID \cite{bray2003mma} and Kalign \cite{lassmann2005kaa}
global sequence alignment programs available for download at \cite{mavid} and \cite{kalign}.
We also experimented with STRETCHER \cite{stretcher},
lagan \cite{brudno2003lam} and CLUSTALW 2 \cite{thompson22gci}, and observed 
similar results. We have made no efforts here to optimize the scoring parameters of the 
algorithms used and have only used the defaults.\pb

For DNA string compression we utilized the {\it expert model} (XM) DNA compression algorithm 
\cite{cao2007ssa}.  For compression of the translation strings we used lpaq1 \cite{mahoney}. Using the lpaq1 was not crucial, 
with the standard LINUX tools gzip and bzip2 producing similar results. For DNA we also explored
GenCompress \cite{li2001ibs} and bzip2. Both showed markedly inferior results to XM 
(see supplementary information).\pb

The complete mtDNA sequences used in our analysis were downloaded from \cite{ncbi}. They
included 220 mammals, 25 non-mammalian vertebrates, and 20 invertebrates.

\section*{Results}

In Figure~\ref{fig:mavid_PG_vs_XM} we compare two  MI estimates for pairs of species 
from various groups of animals. The first estimate is obtained using the MAVID 
alignment tool \cite{bray2003mma} followed by compression, while the second 
is obtained by compression alone \cite{li2001ibs,li2004,cili2005}, without using any 
alignment algorithm. The latter estimate is made by comparing the size of the 
compressed concatenation $AB$ to the sum of the sizes of the compressed individual files,
\begin{equation}
   I(A;B)_{\rm compr} = {\rm len}[{\rm XM}(A)] + {\rm len}[{\rm XM}(B)] - {\rm len}[{\rm XM}(AB)].
\end{equation}
Although it is not possible to prove that $I_{\rm compr}$ or $I_{\rm align}$ are lower bounds 
for the true MI, generally it is expected that both $I_{\rm compr}$ and $I_{\rm align}$ 
underestimate the true MI.\pb

In Figure~\ref{fig:mavid_PG_vs_XM} we find that both MI estimates are
approximately equal, despite the fact that alignment algorithms and compression
algorithms follow drastically different routes.  Points above the diagonal
indicate that concatenation and compression -- using the XM algorithm --
produced a better estimate of MI, while points below indicate that MAVID
alignment followed by compression of its translation string produced a better
estimate.  Different results are found by compressing with compression
algorithms other than XM (see supplementary material).  In that case a vast
majority of the points fall far below the diagonal. The
invertebrate-invertebrate pairs far above the diagonal in
Figure~\ref{fig:mavid_PG_vs_XM} correspond  to pairs of species where the
individual genes are  similar, but their ordering is changed. In that case a
compression algorithm is superior to a global alignment algorithm, since it is
not affected by shuffling the open reading frames (ORFs). Most negative
estimates for  MI seen in Figure~\ref{fig:mavid_PG_vs_XM} represent cases where
shuffling the ORFs prevented reasonable global alignments. Results could have
been improved in such cases by masking part of the genome, but we have not
tried this.\pb

\begin{figure}
\centering
\includegraphics[width=0.7\textwidth]{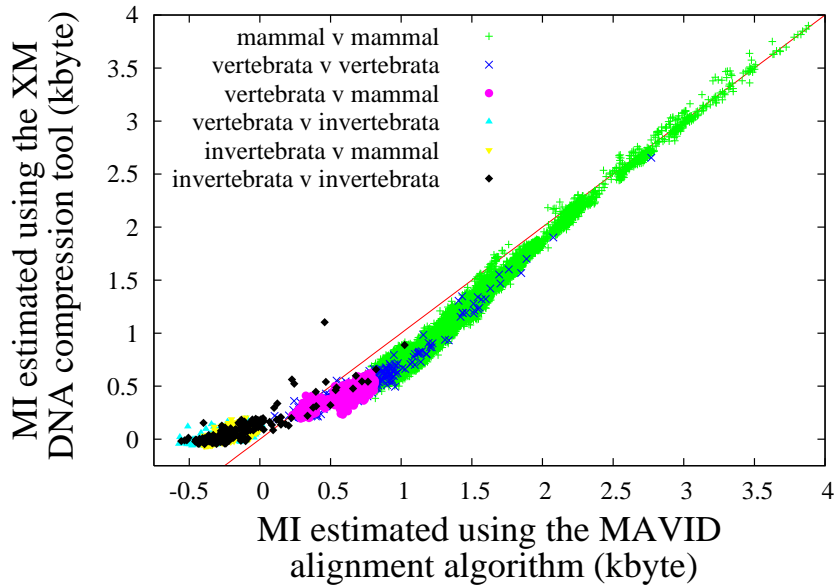}
\caption{Scatter plot of MI estimates for complete mitochondrial DNA between pairs of
  species: $I_{\rm compr}$ using XM \cite{cao2007ssa} vs. $I_{\rm align}$ using 
  MAVID \cite{bray2003mma} followed by compression. Note that the two estimates
  generally agree and fall on the diagonal, while in some cases one method does better 
  than the other as explained in the text. Here and in Figure~\ref{fig:kalign_PG_vs_mavid_PG} 
  ``vertebrata'' means non-mammalian vertebrata. See Tools for a breakdown on the 
  number of mammals, vertebrata and invertebrata. \label{fig:mavid_PG_vs_XM}}
\end{figure}

\begin{figure}
\centering
\includegraphics[width=0.7\textwidth]{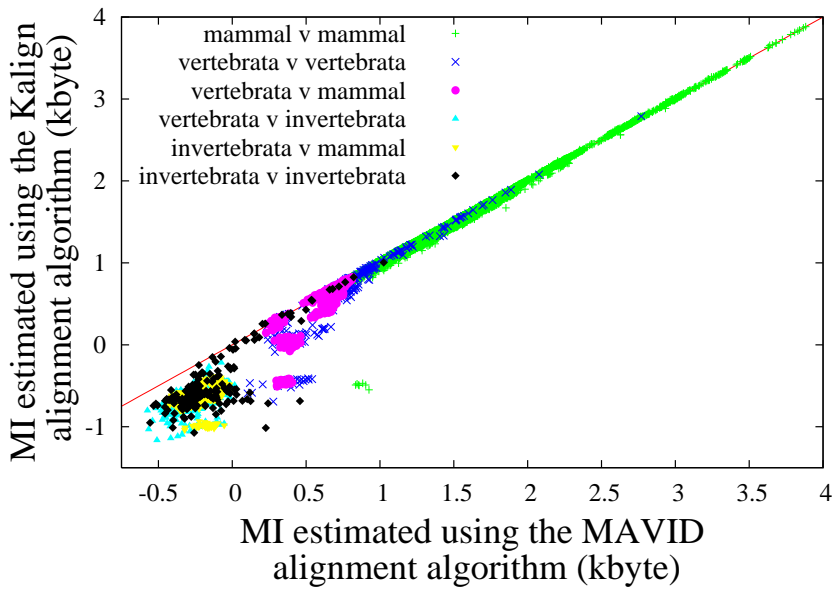}
\caption{Scatter plot comparing alignment based MI estimates for the same pairs of species
  as in Figure~\ref{fig:mavid_PG_vs_XM}: Kalign \cite{lassmann2005kaa} vs. 
  MAVID \cite{bray2003mma}. Points on the diagonal indicate agreement between the two estimates. 
  These data were generated using the default scoring parameters. 
  Therefore, the plot represents a proof of principle rather than a definitive statement
  about the quality of the two alignment algorithms shown. \label{fig:kalign_PG_vs_mavid_PG}
}
\end{figure}

MI estimates obtained using other global  alignment algorithms are  similar to those obtained 
with MAVID; an example is shown in Figure~\ref{fig:kalign_PG_vs_mavid_PG}.  Since neither 
scoring scheme was optimized to obtain this data, we do not consider this figure to indicate
which of the two alignment algorithms is better. Rather, it represents a proof of principle 
that our method can be used to identify strengths and weakness of different alignment algorithms 
and evaluate objectively the similarity of any sequence alignment.

\section*{Discussion}

\noindent Several generalizations and improvements are feasible and are listed below: 

(1) Use more efficient encodings of the translation string. For
instance, we only used the letters $A'_i$ and $T_{B|A,i}$ to get
$B'_i$, but one could also use e.g.  $A'_{i-1}, B'_{i-1}$, and/or
$T_{B|A,i-1}$.

(2) Use local alignments instead of global ones. In a local alignment
between sequences $A$ and $B$, large parts of $B$ are not aligned with
$A$ at all and are encoded without reference to $A$. Only the aligned
parts give information from $A$ that can be used to recover $B$.
Before making the jump from global to local alignments, an
intermediate step would be  a ``glocal'' alignment tool such as
shuffle-lagan (``slagan'') of \cite{brudno2003gaf}. 

(3) Construct objective measures based on information theory for the
quality of multiple alignments.  A straight-forward measure is the
information about sequence $C$ obtained from aligning it
simultaneously with $A$ and $B$. Assume e.g. that the sequences $A$
and $B$ are much more similar to each other than either $A$ and $C$ or $B$ and $C$,
as for human, chimpanzee, and chicken.  In order to measure the MI between
chicken and the primates, one could first align $A$ and $B$ and then align,
in a second step, $C$ to the fixed alignment $(A,B)$.

\section*{Conclusions}
By showing that mutual informations between two sequences can be easily 
estimated from alignments, we have established a direct link between 
sequence alignment and Kolmogorov information theory. Technically, we 
have dealt only with pairwise global alignment, but at least the basic 
concepts should have much wider applicability. We hope that our work will 
be important both for the conventional (alignment-based) approach to 
sequence comparison and for the more recent approach based on 
compression and concatenation based on Kolmogorov theory.

The accuracy of MI estimates based entirely on compression and
concatenation depends crucially on the quality of the compression
algorithm (see Figures~S1,~S2). Indeed, Figure~2 shows that alignment based 
estimates can be superior to those based on compression alone, but 
it also shows in other cases the latter to be superior. It is an open 
question whether alignment-free algorithms for sequence comparison 
\cite{vinga2003} will become more widely used, will eventually displace
alignment-based algorithms, or whether both approaches will merge into a unified
approach.  In any case, tools to compare the successes and failures of either 
approach will be crucial.

\section*{Acknowledgements}
  \ifthenelse{\boolean{publ}}{\small}{}
This research was supported by funds from NSERC, iCORE, and Alberta Advanced Education \& Technology.

{\ifthenelse{\boolean{publ}}{\footnotesize}{\small}
 \bibliographystyle{bmc_article}  
\bibliography{seqcomp_bioinfo_v2} }     

\end{bmcformat}
\end{document}


{\bf \Large Supplementary Material}

\vskip .8cm

\noindent {\bf Blasting and Zipping: Sequence Alignment and Mutual Information}

\vskip .4cm

O. Penner, P. Grassberger, and M. Paczuski

\vskip .8cm

1) Mutual information (MI) provides an absolute and objective measure of the similarity between
any two sequences $A$ and $B$ built from a finite alphabet. The main result of our paper is that MI can be reliably estimated 
from alignments in general, and from {\it global} alignments in particular. This is done by first using 
an alignment between $A$ and $B$ to construct two `translation strings', $T_{B|A}$ and $T_{A|B}$, which allow $B$ to be uniquely reconstructed from $A$, and $A$ from $B$, respectively. After compression, the lengths of these 
strings give estimates of the conditional algorithmic informations $K(B|A)$ and $K(A|B)$. Finally, the 
(algorithmic) MI is estimated by means of the general relations \cite{cover}
\be
    I(A;B) = K(A)-K(A|B), \quad I(B;A) = K(B)-K(B|A),
\ee
where $K(A)$ and $K(B)$ are the algorithmic informations (also called ``Kolmogorov-Chaitin complexities") of 
the sequences $A$ and $B$. These are estimated by the lengths of compressed versions of $A$ and $B$. 
A central result of algorithmic information theory is that $I(A;B) = I(B;A)$ up 
to terms $O(\log N)$, where $N$ is the length of the concatenation $AB$.\cite{cover}

The mutual informations estimated this way, denoted $I_{\rm align}$ (see Eq.~(3)), can be compared to estimates of MI obtained 
without using any alignment. The latter is obtained by comparing the combined lengths of the compressed versions of 
$A$ and $B$ to the length of a compressed version of the concatenation $AB$.  We denote this $I_{\rm compr}$ and it
is given by Eq.~(4). Our main result in the paper is that both estimates, despite being 
independent and following different strategies, yield very similar results for mitochondrial DNA (mtDNA)
of vertebrates. More precisely, they give practically {\it identical} estimates for species within the same 
family. $I_{\rm align}$ is typically slightly larger for species in the same class but in different families,
while $I_{\rm compr}$ is larger for species in different phyla, a case where global alignment algorithms break down.

2) The estimate $I_{\rm compr}$ is the standard estimate for the MI between two strings, and has been used recently 
in a large number of biological and non-biological problems \cite{li2001ibs,li2004,cili2005,kraskov2005hcu,kraskov2008}. 
Most of the work done with $I_{\rm compr}$ has focussed on the clustering of sequences and building phylogenetic trees, however it seems that this activity has 
met with considerable skepticism in the biological community. One reason is that it appears that no biological 
knowledge is incorporated by estimating $I_{\rm compr}$.  This is in stark contrast to the substantial amount of detailed 
knowledge that has gone into the construction of phylogenies based on alignment methods. Another reason is 
that even the order of magnitude of $I_{\rm compr}$ depends on the quality of the compression 
algorithms used. While two compression algorithms can be easily compared via the lengths of the compressed 
files they produce, it is practically impossible to judge the merits of a compression algorithm on an absolute 
scale. Most non-trivial sequences like DNA, proteins, music, or written language show very long and structurally 
complex long-range correlations. To obtain good estimates of $I_{\rm compr}$ it is crucial that these long-range
effects are correctly taken into account; however, the precise structure and the related effect on the compressibility
are usually not known. 

\begin{figure}
\centering
\includegraphics[width=0.46\textwidth]{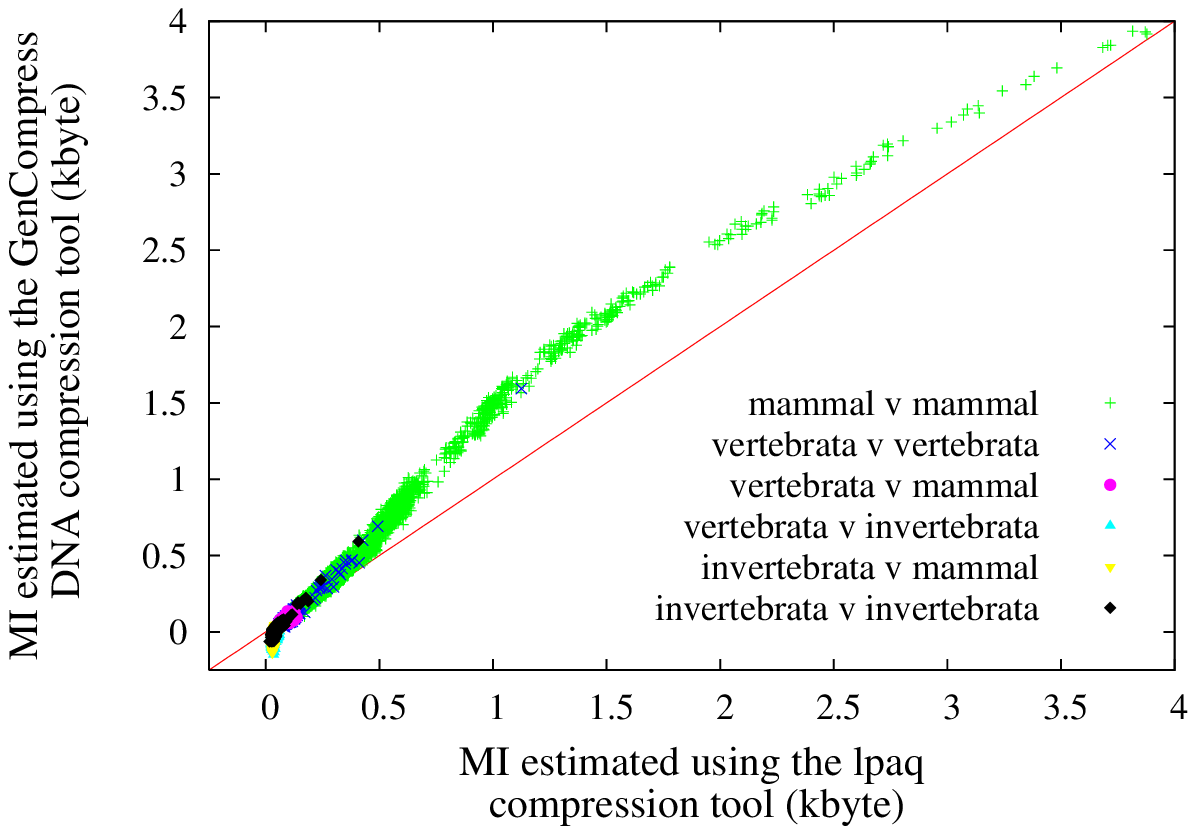}
\caption{Scatter plot of compression-based MI estimates for complete mitochondrial DNA between pairs of
species: $I_{\rm compr}$ using GenCompress~\cite{li2001ibs} versus $I_{\rm compr}$ using lpaq1~\cite{mahoney}.
As in Fig.~2 and 3, ``vertebrata'' indicates non-mammalian vertebrata. }
\end{figure}

\begin{figure}
\centering
\includegraphics[width=0.46\textwidth]{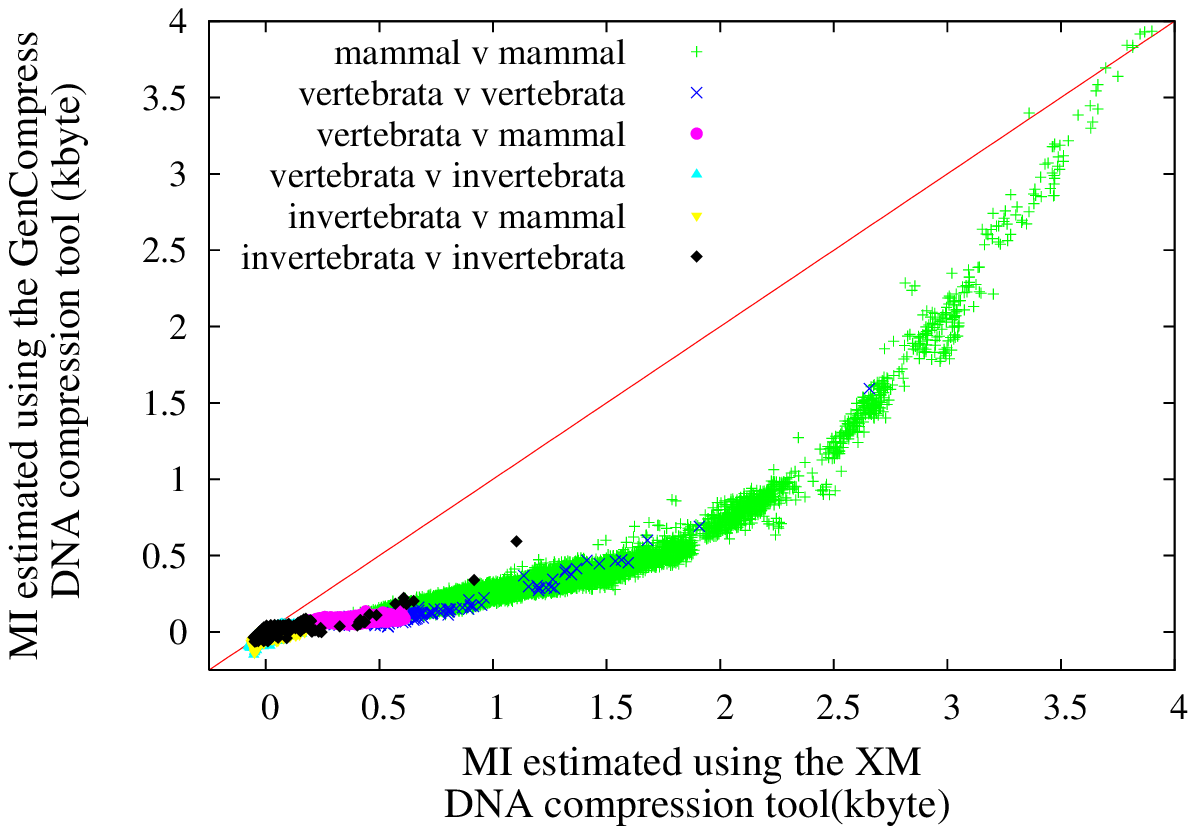}
\caption{Similar to Fig.~S1, but $I_{\rm compr}$ using XM~\cite{cao2007ssa} versus $I_{\rm compr}$ using GenCompress.}
\end{figure}

It is generally accepted that up to approximately five years ago the most advanced general purpose compression algorithms were not very efficient 
for DNA \cite{grumbach,gencomp}.  However, there has been substantial progress during the last five years in 
this field \cite{mahoney}. Nevertheless, we found that none of the recent general purpose algorithms we checked 
(lpaq1, durilca, bbb, dark \cite{mahoney}) gave better compression 
rates for DNA than GenCompress~{\cite{li2001ibs}}, one of the best public domain DNA compression algorithm known 
prior to 2007. A typical plot ($I_{\rm compr}$ using lpaq1 versus $I_{\rm compr}$ using GenCompress for mtDNA) is 
shown in Fig.~S1.

On the other hand, with the development of XM~\cite{cao2007ssa} there has been important progress in biosequence (DNA, proteins) specific 
compression algorithms. According to the authors of Ref.~\cite{cao2007ssa}, the improvement 
brought by XM over previous DNA compression algorithms (such as GenCompress) is only a few per cent, when single 
sequences are considered. However, we found that the improvement in estimates of $I_{\rm compr}$ is much larger, 
presumably because long range correlations play a much bigger role there. 
Results for $I_{\rm compr}$ using GenCompress versus $I_{\rm compr}$ using XM are shown in Fig.~S2. We observe vast differences, except for 
very closely related species, where both compression algorithms are able to detect the strong similarity. For species in different families, XM gives typically three to five times larger 
MIs than GenCompress. Note that we do not have a rigorous proof that larger values of $I_{\rm compr}$ 
are better. The difficulty is that $I_{\rm compr}$ is the difference between terms which are all overestimated.
But the negative term, corresponding to $K(AB)$, is most difficult to estimate because $AB$ is much longer than $A$ or $B$.  Thus $K(AB)$ is
most likely to be strongly overestimated. As such, it follows that larger values of $I_{\rm compr}$ indicate 
improved treatment of long-range interdependencies and better estimates of MI.

The strong dependence on compression algorithm observed in Fig.~S2 seems to justify skepticism against the use of 
$I_{\rm compr}$. This is contradicted, however, by the fact that the values of $I_{\rm compr}$ obtained with XM
are in very good general agreement with the values of $I_{\rm align}$, as shown in Fig.~2. The latter suggests 
that it should be possible to improve $I_{\rm compr}$ further by a factor $\approx 2$ for species in different phyla,
but probably not by more.

\begin{figure}
\centering
\includegraphics[width=0.35\textwidth,angle=270]{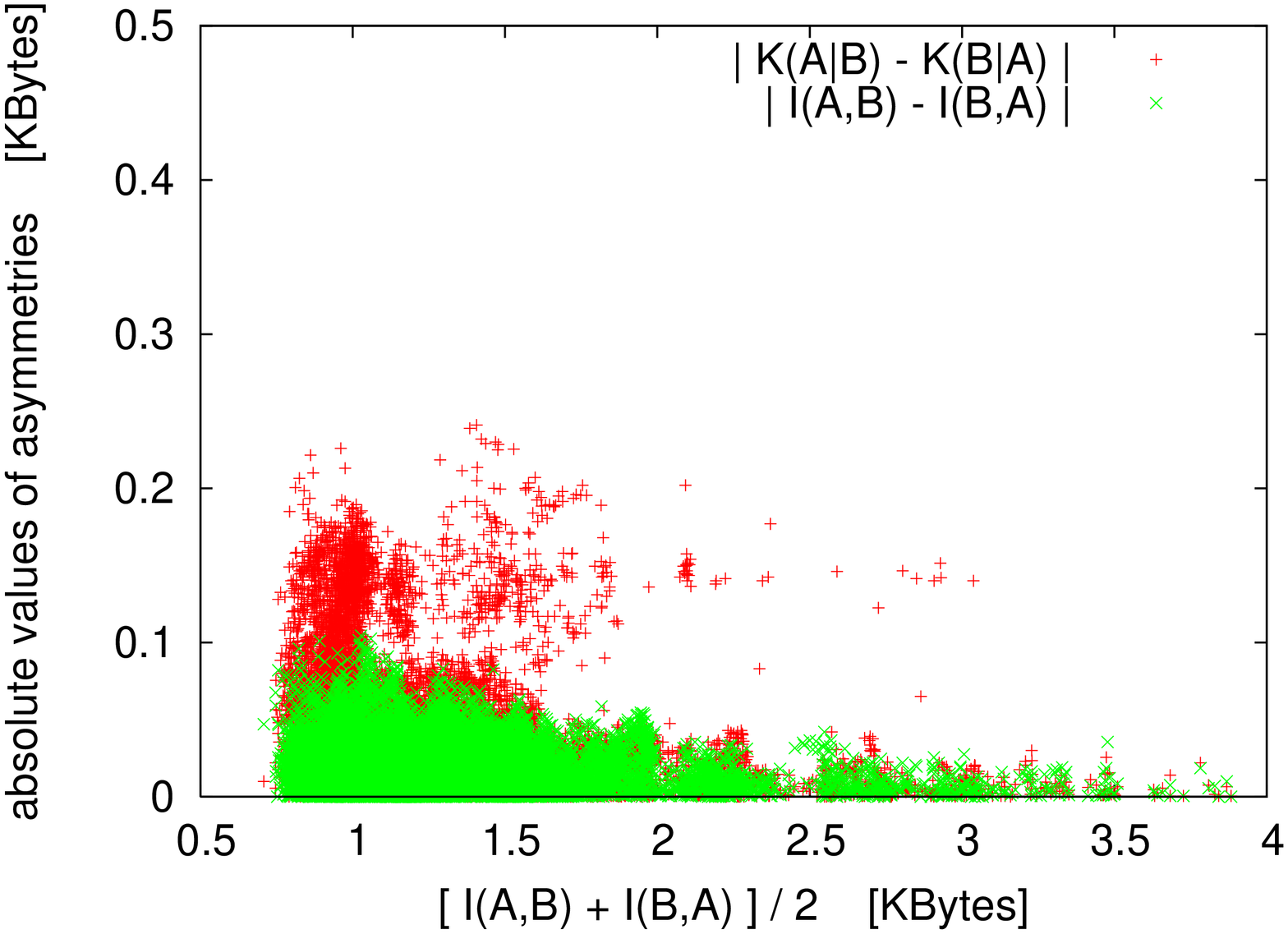}
\caption{Scatter plot of asymmetries of $I_{\rm align}$ and of conditional informations, both versus 
$I_{\rm align}$. Pairs of species are drawn only from mammals.}
\end{figure}

3) An important aspect of our treatment is that the translation strings $T_{B|A}$ and $T_{A|B}$ are different. 
Thus the conditional informations $K(B|A)$ and $K(A|B)$ are also different. This is in stark contrast to 
the notion of edit distances~{\cite{navarro2001}}, where one typically wishes to define a symmetric distance measure directly via the number 
and costs of edit operations. In our treatment, on the other hand, the symmetric distance is obtained in further steps, 
by first going over to MIs and then, if one wishes, by deriving universal compression distances from the MI 
\cite{li2004}. Although this, at first, appears more complicated, our method has the advantage of providing a direct link
with information theory.

A crucial numerical requirement for our formalism is that the estimates $I_{\rm align}$ should be symmetric
under the exchange of the sequences within terms of $O(\log N)$, $I(A;B)_{\rm align} \approx I(B;A)_{\rm align}$.
Any strong violation of this symmetry would indicate that either the construction of the translation string is 
not optimal, or that the compression algorithm used in Eq.~(3) is far from perfect. In contrast 
it is not required that $K(B|A)$ is symmetric. To show that $I(A;B)_{\rm align}$ is more symmetric than
$K(B|A)$, we plot in Fig.~S3 the differences $| I(A;B)_{\rm align}-I(B;A)_{\rm align}|$ and 
$|K(B|A) - K(A|B)|$ against $I(A;B)_{\rm align}$. For Fig.~S3 we used only mammalian mtDNA because the 
estimates $I(A;B)_{\rm align}$ for species in different classes are too uncertain for a meaningful analysis.
We see that there are no problems at all for closely related species as in such cases, both $I(A;B)_{\rm align}$ and 
$K(B|A)$ are symmetric. For more distant species, both are still symmetric for the majority of pairs, but there 
are also numerous outliers where $K(B|A)$ is strongly asymmetric. In all those cases the asymmetry of 
$I(A;B)_{\rm align}$ is significantly smaller than that of $K(B|A)$.

4) In the paper we have argued qualitatively how the symmetry of $I(A;B)_{\rm align}$ is compatible, in view of 
Eq.~(3), with even very asymmetric values of $K(B|A)$. Here we shall discuss some extreme cases quantitatively and 
exactly.

\begin{figure}
\centering
\includegraphics[width=0.46\textwidth]{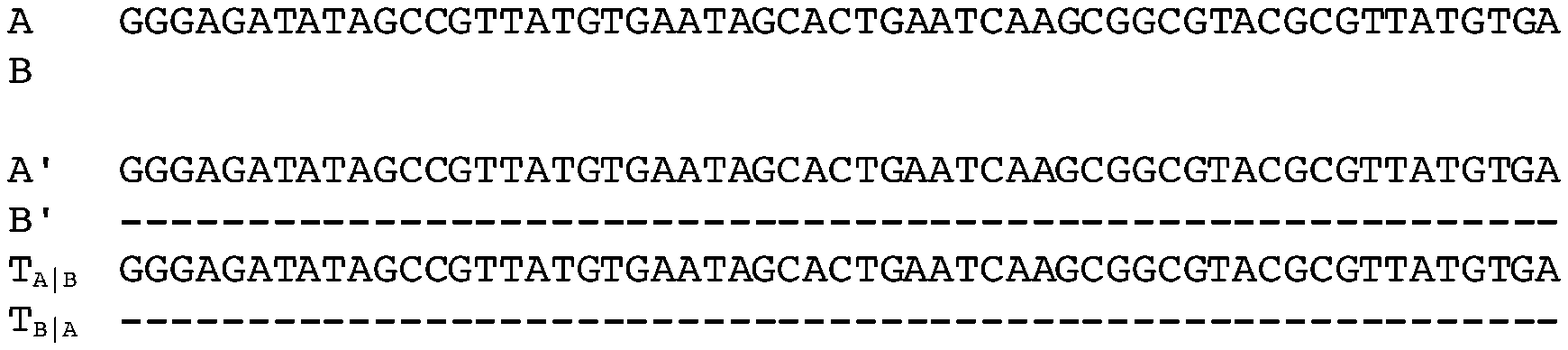}
\caption{Alignment and translation strings for comparing a random sequence with an empty sequence. Here, sequence
$A$ is random and of length $n$, while $B$ is empty. As explained in the text, the estimated MIs $I(B;A)_{\rm align}$
and $I(A;B)_{\rm align}$ agree with the expected results to within terms of order ${\cal O}(\log n)$.}
\end{figure}

\begin{figure}
\centering
\includegraphics[width=0.46\textwidth]{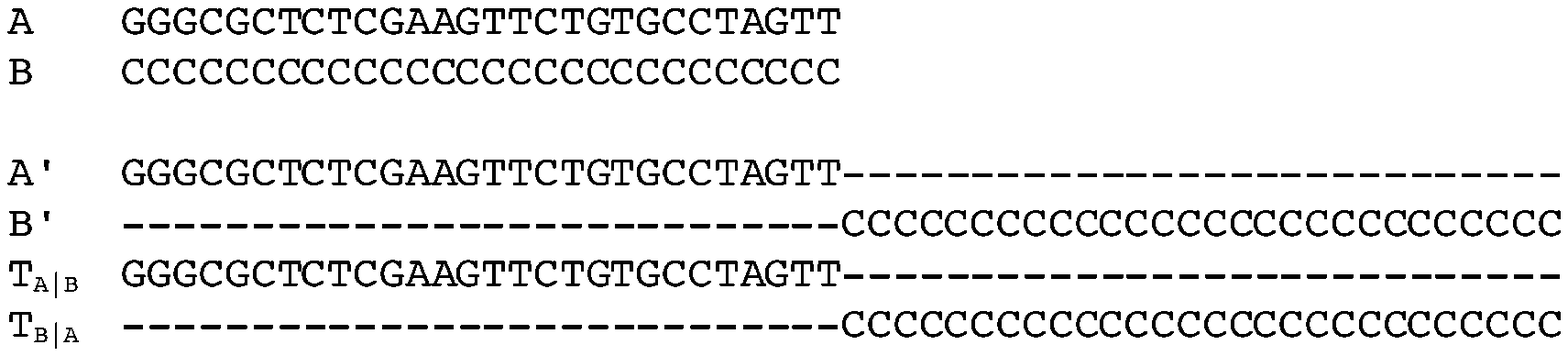}
\caption{Alignment and translation strings for comparing a random sequence with a sequence composed of only one 
letter. Here, sequence $A$ is random of length $n$, while $B$ is a string of length $n$ consisting only of ``C". As explained 
in the text, the estimated MIs $I(B;A)_{\rm align}$
and $I(A;B)_{\rm align}$ agree with the expected results to within terms of order ${\cal O}(\log n)$.}
\end{figure}

\begin{figure}
\centering
\includegraphics[width=0.46\textwidth]{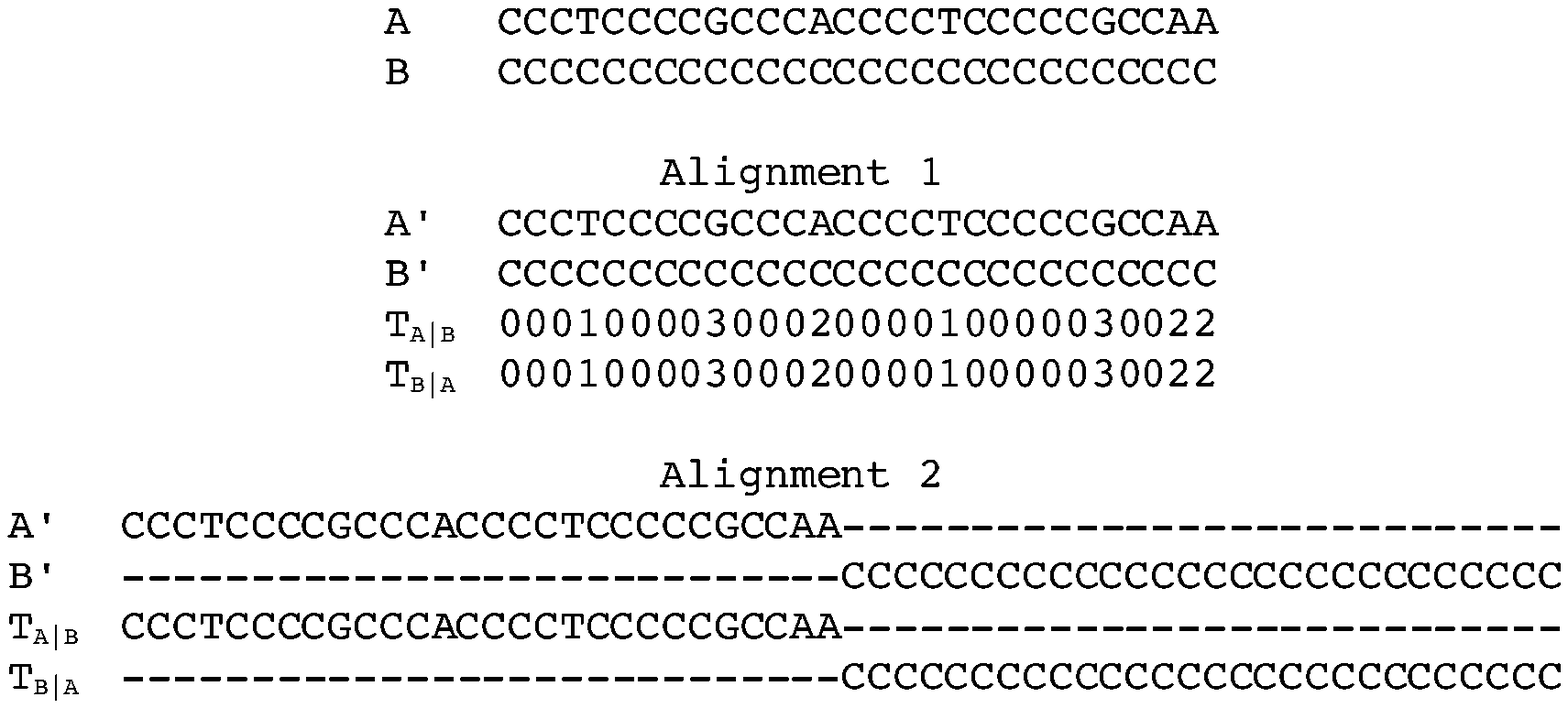}
\caption{Two alternative alignments between a biased random sequence and a sequence composed of only one letter. 
See the text for details.} 
\end{figure}

a) Assume that the string $A$ is a random string over the alphabet $\{A,C,G,T\}$ of length $n$, and the string $B$ 
is the empty string. Then the optimal alignment is as shown in Fig.~S4. In order to specify $T_{B|A}$ one must encode 
the letter ``-" and the number $n$ of repetitions, giving $K(B|A) \approx \log_2n $ bits, while $K(B)=0$. On the other hand, $K(A|B) 
=K(A)$. These give $I(B;A)_{\rm align} \approx -\log_2n $ bits and $I(A;B)_{\rm align} ={\cal O}(1)$. The 
reason why $I(A;B)_{\rm align}$ is not exactly zero is that the two terms on the r.h.s. of Eq.~(3) are 
compressed by means of different algorithms. Both have to be given verbatim, so the only difference 
between the two term in Eq.~(3) is the difference between the overheads in lpaq1 and XM.

In summary, $I(B;A)_{\rm align}-I(A;B)_{\rm align} = O(\log n)$, as expected on general grounds and for 
the difference between the exact MI values.

b) Assume that $A$ is a random string over the alphabet $\{A,C,G,T\}$ of length $n$, and $B$ a string of the same
length consisting of a single letter, say ``C". The optimal alignment for this case is shown in Fig.~S5.
Now $T_{B|A}$ consists of $n$ hyphens followed by $n$ ``C"s, which gives $I(B;A)_{\rm align} \approx -\log_2 n$ bits.
Similarly, $T_{A|B} = A$, followed by $n$ hyphens, so that also $I(A;B)_{\rm align} \approx -\log_2 n$. Thus,
for Fig.~S5 the difference $I(B;A)_{\rm align}-I(A;B)_{\rm align}$ is again as expected for the exact MI values.

c) Finally, we can consider a situation similar to case b), but with $A$ not fully random. Instead, we
assume that $A$ is iid. with ${\rm prob}(A_i = {\rm C})\gg {\rm prob}(A_i ={\rm A})= {\rm prob}(A_i = {\rm G})= 
{\rm prob}(A_i = {\rm T})$. In this case one might be inclined intuitively to prefer alignment \#1 of Fig.~S6
over alignment \#2. But for alignment \#1 both $T_{B|A}$ and $T_{A|B}$ would be equal to $A$, up to a complexity
preserving transformation A$\to$1, C$\to$0, G$\to$3, and T$\to$2. Thus $I(B;A)_{\rm align} \propto n$, while 
$I(A;B)_{\rm align} = {\cal O}(1)$. On the other hand, for alignment \#2 both $I(B;A)_{\rm align}$ and 
$I(A;B)_{\rm align}$ are the same as in Fig.~S5. Thus alignment \#2 is the optimal alignment, although 
one might have preferred alignment \#1 intuitively.

5) Up to now, we have assumed that the two DNA files contain only the four letters ``A", ``C", ``G", and ``T". In
reality, the data banks allow also for ``wild card" letters indicating ambiguities: ``N" for any nucleotide,
``R" for a purine, ``Y" for a pyrimidine, etc. Whenever either of the two sequences contains
such a wild card character, we put $T_{B|A,i} = B_i$, i.e. the letter in the target string is copied verbatim.
This will in some cases slightly overestimate the conditional information $K(B|A)$. But such over estimations 
are expected to occur anyhow, and in the data base we used, wild cards are sufficiently rare to have very 
little effect.

6) In the case of proteins (i.e. amino acid sequences) we have an alphabet of 20 letters. This makes 
the analysis a bit more lengthy, although it is basically the same as for DNA. The translation strings now
contains 41 characters: Twenty letters for specifying insertions, one hyphen for indicating a deletion,
and 20 numbers for indicating substitutions. For the latter we have some freedom. As for DNA, we could use 
this freedom to encode forward and backward substitutions by the same number. Alternatively, we could 
encode the 20 amino acids by numbers $0\ldots 19$, and encode a substitution $j\to k$ ($j,k=0\ldots 19$)
by the number $k - j\;\;{\rm mod}\;20$. This has the advantage of simplicity, but led in the case of DNA to 
marginally worse results.

7) For local alignments, the output of an alignment algorithm consists of a list of matching regions, 
together with the actual alignments of those regions. These matching regions can in principle overlap.
Thus in order to construct $T_{B|A}$, one first has to select a subset of matching regions ${\cal M}_k$ that do 
not overlap on sequence $B$. Each one of these matches is characterized by its starting points in 
sequences $A$ and $B$ and by the translation string restricted to the matching regions,
\be
   {\cal M}_k \equiv \{n_k^{(A)},n_k^{(B)}, T_{B|A;k}\} .
\ee
The entire translation string consists then of all these pieces of information, separated by uniquely
decodable separators, plus the verbatim description of sequence $B$ in regions where no matches were found or used.
Indeed, if one has the latter verbatim description and all translation strings $T_{B|A;k}$ for 
the matching regions, one can recover the points $n_k^{(B)}$ and does not have to include them in 
$T_{B|A}$ explicitly.

\begin{figure}
\centering
\includegraphics[width=0.46\textwidth]{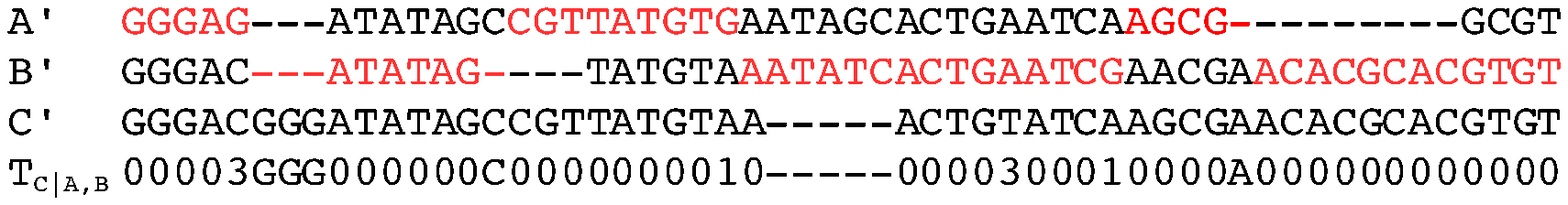}
\caption{Alignment of a third sequence $C$ to an already existing alignment between $A$ and $B$. The 
translation string $T_{C|AB}$ for reconstructing $C$ from $A$ and $B$ is obtained by using locally 
one of the strings $A$ and $B$ as ``master strings" and applying the rules described in the main part 
of the paper. The actual master string is printed in red.}
\end{figure}

8) As a first step beyond pairwise alignments and pairwise MIs, we shall discuss the estimation of the MI 
between a single sequence $C$ and a pair $(A,B)$. For example, an application of this could be to estimate
the similarity between mouse and the family of hominids, where the latter is characterised by the two 
species homo and chimpanzee.

Estimating $I(C;(A,B))$ via concatenation and compression alone is easy. One just has to modify Eq.~(4) to
\begin{eqnarray}
   I(C;(A,B))_{\rm compr} & = & {\rm len}[{\rm XM}(AB)]+{\rm len}[{\rm XM}(C)] \nonumber\\
   & & - {\rm len}[{\rm XM}(ABC)]
\end{eqnarray}
where $AB$ and $ABC$ denote the concatenations of $A$ with $B$ and of $AB$ with $C$. This gives
$I(C;(A,B))_{\rm compr} \geq I(C;A)_{\rm compr}$ and $I(C;(A,B))_{\rm compr} \geq I(C;B)_{\rm compr}$,
in agreement with general relations between MIs and with intuition. In contrast, concatenating $A$ and $B$
and then aligning (globally!) $C$ with $AB$ would lead to very poor estimates of $I(C;(A,B))_{\rm align}$.

Instead, one has to use progressive multiple alignments, where one first aligns the two sequences $A$ and $B$,
and then aligns the third sequence $C$ to the fixed alignment of the first two. The central problem is 
to construct a translation string which allows to reconstruct $C$ from the aligned pair $(A,B)$.

One possibility is the following, illustrated in Fig.~S7. We start at position $i=1$ and construct 
$T_{C|AB}$ from left to right, using at each step one of $A$ or $B$ as the ``master sequence".
If $A$ is presently the master sequence, then $T_{C|AB,i} = T_{C|A,i}$, where $T_{C|A,i}$ is as 
described in the main paper. Similarly, if $B$ is the master sequence, then $T_{C|AB,i} = T_{C|B,i}$. 
A sequence ($A$ or $B$) is kept as the master sequence until $C'_i$ disagrees with the character in the master 
sequence {\it and} is identical to the character in the other (non-master) sequence.  At this point 
master and non-master sequences switch their status. In a slightly more sophisticated version, one 
keeps track of the number of `mistakes' made recently by sequences $A$ and $B$, and switches only when
the current master has done significantly worse over the recent past.
We have no proof that either encoding is optimal, but both guarantee at least that the ``better" of the 
two sequences $A$ and $B$ is used as a template to reconstruct $C$.

We have made very preliminary numerical tests showing that the proposals made in points 6) to 8) are 
potentially feasible, but much more complete investigations are needed and will be given in future 
publications.

\bibliography{seqcomp_v7}